# United Nations Basic Space Science Initiative: 2011 Status Report on the International Space Weather Initiative

S. Gadimova and H.J. Haubold, United Nations Office for Outer Space Affairs, Vienna International Centre, PO Box 500, A-1400 Vienna, Austria

D. Danov and K. Georgieva, Solar Terrestrial Influences Laboratory, Bulgarian Academy of Sciences, Block 3, Academician G. Bonchev Str., 1113 Sofia, Bulgaria

G. Maeda and K. Yumoto, Space Environment Research Center, Kyushu University 53, 6-10-1 Hakozaki Higashi-ku Fukuoka, Postal Code 812-8581, Japan

J.M. Davila and N. Gopalswami, Laboratory for Solar and Space Physics, NASA/GSFC, Code 671, Solar Physics Branch, Greenbelt, MD 20771, USA

Abstract. The UBSSI is a long-term effort for the development of astronomy and space science through regional and international cooperation in this field on a worldwide basis, particularly in developing nations. A series of workshops on BSS was held from 1991 to 2004 (India 1991, Costa Rica and Colombia 1992, Nigeria 1993, Egypt 1994, Sri Lanka 1995, Germany 1996, Honduras 1997, Jordan 1999, France 2000, Mauritius 2001, Argentina 2002, and China 2004; http://www.seas.columbia.edu/~ah297/un-esa/) and addressed the status of astronomy in Asia and the Pacific, Latin America and the Caribbean, Africa, and Western Asia. One major recommendation that emanated from these workshops was the establishment of astronomical facilities in developing nations for research and education programmes at the university level. Material for teaching and observing programmes for small optical telescopes were developed or recommended and astronomical telescope facilities have been inaugurated in a number of nations. Such workshops on BSS emphasized the particular importance of astrophysical data systems and the virtual observatory concept for the development of astronomy on a worldwide basis. Pursuant to resolutions of the United Nations Committee on the Peaceful Uses of Outer Space (UNCOPUOS) and its Scientific and Technical Subcommittee, since 2005, these workshops focused on the International Heliophysical Year 2007 (UAE 2005, India 2006, Japan 2007, Bulgaria 2008, Ro Korea 2009; http://www.unoosa.org/oosa/SAP/bss/ihy2007/index.html). Starting in 2010, the workshops focus on the International Space Weather Initiative (ISWI) as recommended in a three-year-work plan as part of the deliberations of UNCOPUOS (http://www.stil.bas.bg/ISWI/). Workshops on the ISWI have been scheduled to be hosted by Egypt in 2010 for Western Asia, Nigeria in 2011 for Africa, and Ecuador in 2012 for Latin America and the Caribbean. Currently, 14 IHY/ISWI instrument arrays with > 600 instruments are operational in 95 countries.

## I. Introduction

### A. Background and objectives

The Third United Nations Conference on the Exploration and Peaceful Uses of Outer Space (UNISPACE III), in particular through its resolution entitled "The Space Millennium: Vienna Declaration on Space and Human Development", recommended that activities of the United Nations Programme on Space Applications should promote collaborative participation among Member States, at both the regional and international levels, in a variety of space science and technology activities, by emphasizing the development and transfer of knowledge and skills to developing countries and countries with economies in transition.[1]

At its fifty-second session, in 2009, the Committee on the Peaceful Uses of Outer Space endorsed the programme of workshops, training courses, symposiums and conferences planned for 2010.[2] Subsequently, the General Assembly, in its resolution 64/86, endorsed the report of the Committee on the work of its fifty-second session.

Pursuant to General Assembly resolution 64/86 and in accordance with the recommendations of UNISPACE III, annual United Nations/National Aeronautics and Space Administration/Japan Aerospace Exploration Agency Workshops on the International Space Weather Initiative have been scheduled to be held prospectively in Egypt (2010) for the region of Western Asia, in Nigeria (2011) for the region of Africa, and in Ecuador (2012) for the region of Latin America and the Caribbean.

Organized by the United Nations, the European Space Agency (ESA), the National Aeronautics and Space Administration (NASA) of the United States of America and the Japan Aerospace Exploration Agency (JAXA), the Workshop was the eighteenth in a series of workshops on basic space science, the International Heliophysical Year 2007 and the International Space Weather Initiative proposed by the Committee on the Peaceful Uses of Outer Space on the basis of discussions of its Scientific and Technical Subcommittee, as reflected in the report of the Subcommittee (A/AC.105/958, paras. 162-173). Previous workshops in the series were hosted by the Governments of the United Arab Emirates in 2005 (A/AC.105/856), India in 2006 (A/AC.105/882), Japan in 2007 (A/AC.105/902), Bulgaria in 2008 (A/AC.105/919) and the Republic of Korea in 2009 (A/AC.105/964).[3] Those workshops were a continuation of the series of workshops on basic space science that were held between 1991 and 2004 and that were hosted by the Governments of India (A/AC.105/489), Costa Rica and Colombia (A/AC.105/530), Nigeria (A/AC.105/560/Add.1), Egypt (A/AC.105/580), Sri Lanka (A/AC.105/640), Germany (A/AC.105/657), Honduras

---

[1] *Report of the Third United Nations Conference on the Exploration and Peaceful Uses of Outer Space, Vienna, 19-30 July 1999* (United Nations publication, Sales No. E.00.I.3), chap. I, resolution 1, sect. I, para. 1 (e)(ii), and chap. II, para. 409 (d)(i).
[2] *Official Records of the General Assembly, Sixty-fourth Session, Supplement No. 20* (A/64/20), para. 82.
[3] Information on the International Heliophysical Year 2007 and the United Nations Basic Space Science Initiative is available on the website of the Office for Outer Space Affairs at www.unoosa.org/oosa/SAP/bss/ihy2007/index.html.

(A/AC.105/682), Jordan (A/AC.105/723), France (A/AC.105/742), Mauritius (A/AC.105/766), Argentina (A/AC.105/784) and China (A/AC.105/829).[4]

The main objective of the Workshops was to provide a forum in which participants could comprehensively review achievements of the International Heliophysical Year 2007 in terms of the deployment of low-cost, ground-based, worldwide space weather instruments and plans for the International Space Weather Initiative and assess recent scientific and technical results in the field of solar-terrestrial interaction.

## B. Workshop Programmes

In general, the Workshops focused on the following topics: national coordination of the International Space Weather Initiative, instrument arrays of the Initiative in operation, distribution of Initiative instruments by countries, and data analysis and modelling.

## C. Attendance

Scientists, engineers and educators from developing and industrialized countries from all economic regions were invited by the United Nations, NASA, JAXA, the International Committee on Global Navigation Satellite Systems (ICG), the Space Environment Research Centre (SERC) of Kyushu University in Fukuoka, Japan, Helwan University and the Space Weather Monitoring Center of Egypt to participate at and contribute to the Workshop. Workshop participants, who held positions at universities, research institutions, national space agencies and international organizations, were involved in implementing activities of the International Space Weather Initiative covered by the Workshop. Participants were selected on the basis of their scientific, engineering and educational backgrounds and their experience in implementing programmes and projects in which the Initiative played a leading role. The preparations for the Workshop were carried out by an international scientific organizing committee, a national advisory committee and a local organizing committee.

A list of designated national and area coordinators of the International Space Weather Initiative, status December 2010, is provided in annex I to the present document. A table summarizing the type and number of International Space Weather Initiative instruments by country or area, status December 2010, is provided in annex II.

## II. International Space Weather Initiative Instrument Arrays in Operation

### 1. African Global Positioning System Receivers for Equatorial Electrodynamics Studies

It was recalled that the African Global Positioning System Receivers for Equatorial Electrodynamics Studies (AGREES) instrument array was deployed to:

---

[4] Details of all the workshops of the United Nations Basic Space Science Initiative organized jointly with the European Space Agency have been made available on the Internet at www.seas.columbia.edu/~ah297/un-esa.

(a) To understand the unique structures of the equatorial ionosphere that had been reported from satellite observation data in the African region, which data had not been confirmed, validated or studied in detail by observations from the ground owing to a lack of suitable ground-based instrumentation in the region;

(b) To monitor and understand the processes governing electrodynamics and plasma production and loss in the lower and middle latitudes as a function of local time, season and magnetic activity;

(c) To estimate the contribution of ionospheric and plasma spherical irregularities and their effect on global navigation satellite systems (GNSS) and communications systems in the African region, where significant signal degradation (scintillation) had become a challenging problem.

2. **African Dual Frequency Global Positioning System Network**

It was noted that the Global Positioning System (GPS) consisted of a minimum of 24 satellites orbiting the Earth at an altitude of approximately 20,000 km. Each satellite transmitted a radio-wave signal to GPS receivers. By determining the time that the GPS signal reached a GPS receiver, one calculated the distance to the satellite in order to determine the exact position of the GPS receiver on Earth. Different errors in the determination of the distance between satellite and GPS receiver were introduced while the signal traversed the ionosphere and troposphere. The analysis of the satellite signal errors led to the determination of geophysical parameters such as the total electron content in the ionosphere or atmospheric water vapour distribution in the troposphere. The African Dual Frequency Global Positioning System Network (GPS-Africa) instrument array consisted of a number of different networks of GPS receivers: the International GPS Service (IGS), the Analyse multidisciplinaire de la mousson africaine (AMMA), the Scintillation Network Decision Aid (SCINDA) and AGREES.

3. **African Meridian B-field Education and Research**

It was observed that the African Meridian B-field Education and Research (AMBER) instrument array was deployed to: (a) monitor the electrodynamics that governed the motion of plasma in the lower and middle latitudes as a function of local time, season and magnetic activity; (b) understand ultra-low-frequency pulsation strength into low- and mid-latitudes and its connection with equatorial electrojets and the auroral electrojet index; and (c) support studies about the effects of Pc5 ultra-low-frequency waves on the mega electron volt electron population in the inner parts of the Van Allen radiation belts.

In addition, to cover the largest land-based gap in global magnetometer coverage, the AMBER instrument array addressed two fundamental areas of space physics: (a) the processes governing the electrodynamics of the equatorial ionosphere as a function of latitude (or L-shell), local time, longitude, magnetic activity and season; and (b) ultra-low-frequency pulsation strength and its connection with equatorial electrojet strength at low- and mid-latitude regions.

Space-based observations showed unique equatorial ionospheric structures in the African region, although those had not been confirmed by ground-based observations, owing to a lack of ground-based instruments in the region. The AMBER magnetometer array, in partnership with GPS receiver arrays (GPS-

Africa, SCINDA and CIDR), would allow the understanding of the electrodynamics that governed equatorial ionospheric motions.

4. **Atmospheric Weather Electromagnetic System for Observation Modeling and Education and the sudden ionospheric disturbance monitor**

   It was recalled that the Atmospheric Weather Electromagnetic System for Observation Modeling and Education (AWESOME) and the sudden ionospheric disturbance monitor instrument arrays consisted of extreme low frequency and very low frequency receivers recording radio signals between 300 Hz and 50 kHz. Monitoring the strength of those signals served as an ionospheric diagnostic tool, since the propagation of the radio signals from transmitter to receiver relied on the conditions of the lower ionosphere.

   The AWESOME instruments recorded a number of single-frequency radio stations and also recorded broadband natural radio signals, such as those which were emitted by lightning and wave-particle interactions in the Earth's magnetosphere. AWESOME monitored the amplitude and phase of very low frequency transmitter signals with 50 Hz time resolution and allowed the entire radio spectrum between 300 Hz and 50 kHz to detect natural signals such as those coming from sferics, whistlers, chorus and hiss. The sudden ionospheric disturbance monitor instruments were a simpler version of the AWESOME instruments for educational purposes and recorded primarily single-frequency stations with an amplitude of very low frequency transmitter signals with 0.2 Hz time resolution.

5. **Compound Astronomical Low-cost Low-frequency Instrument for Spectroscopy and Transportable Observatory**

   It was noted that the Compound Astronomical Low-cost Low-frequency Instrument for Spectroscopy and Transportable Observatory (CALLISTO) spectrometer was a heterodyne receiver. It operated between 45 and 870 MHz, using modern commercially available broadband cable-TV tuners with a frequency resolution of 62.5 kHz. The data recorded by the CALLISTO instrument array were flexible image transport system (FITS) files with up to 400 frequencies per weep. The data were transferred via a R232 cable to a computer and saved locally. Time resolution was of the order of 0.25 second, depending on the number of channels. The integration time was 1 millisecond and the radiometric bandwidth about 300 kHz. The overall dynamic range was larger than 50 decibels.

6. **Continuous H-alpha Imaging Network**

   It was observed that, in order to understand and predict the space weather situation, it was critical to observe erupting phenomena on the solar surface that were initial boundary conditions for all processes. The Continuous H-alpha Imaging Network (CHAIN) instrument array is an observational network with ground-based solar flare monitoring telescopes.

7. **Coherent Ionospheric Doppler Receiver**

   It was recalled that the Coherent Ionospheric Doppler Receiver (CIDR) instrument array consisted of systems of ultra high frequency/very high frequency radio receivers, a control computer and two antennas (one for CIDR and one for GPS). CIDR data were used to tomographically reconstruct the ionosphere along the respective satellite track. Depending on the number of ground installations (no less than three) and the baseline, the tomography could reveal the large-scale structure of the ionosphere, medium-sized structures such as plumes and patches and very fine structures, using a short baseline configuration. In addition, CIDR data were used as an input to data assimilation models for reconstructing the ionosphere on a global or local scale.

8. **Global Muon Detector Network**

    It was noted that the Global Muon Detector Network (GMDN) was a network of multidirectional muon telescopes distributed on three different continents, covering a global range of asymptotic telescope views. As a test case, using GMDN data, it was possible to observe a cosmic ray precursor for the magnetic storm that had occurred in December 2006.

9. **Magnetic Data Acquisition System**

    It was observed that the Magnetic Data Acquisition System (MAGDAS) was deployed for space weather studies during the period 2005-2008, overlapping with the development of the United Nations Basic Space Science Initiative and International Heliophysical Year campaign. MAGDAS aided the study of the dynamics of geospace plasma changes during magnetic storms and aurora substorms, the electromagnetic response of the ionosphere-magnetosphere to various solar wind changes and the penetration and propagation mechanisms of DP2-ULF range disturbances from the solar wind region into the equatorial ionosphere. MAGDAS conducted real-time monitoring and modelling of the global three-dimensional current system and the ambient plasma density for understanding changes in the electromagnetic and plasma environment in the geospace.

10. **Optical Mesosphere Thermosphere Imager**

    It was recalled that the Optical Mesosphere Thermosphere Imager (OMTI) instrument array observed the Earth's upper atmosphere through nocturnal airglow emissions from oxygen and hydroxyl in the mesopause region (at analtitude 80-100 km) and from oxygen in the thermosphere/ionosphere (at altitude of 200-300 km). OMTI consisted of all-sky cooled charge-coupled device imagers, Fabry-Perot interferometers, meridian scanning photometers and airglow temperature photometers, in order to measure two-dimensional images of upper atmospheric disturbances and their Doppler wind and temperature.

11. **Remote Equatorial Nighttime Observatory for Ionospheric Regions**

    It was noted that Remote Equatorial Nighttime Observatory for Ionospheric Regions (RENOIR) stations operated in order to improve understanding of the variability in the night-time ionosphere and the effects of that variability on critical satellite navigation and communication systems. RENOIR instruments were dedicated to studying the equatorial/low-latitude ionosphere/thermosphere system, its response to storms and the irregularities that appeared on a daily basis. A RENOIR station consisted of the following: (a) one wide-field ionospheric imaging system; (b) two miniaturized Fabry-Perot interferometers; (c) a dual-frequency GPS receiver; and (d) an array of five single-frequency GPS scintillation monitors. The array of single-frequency GPS scintillation monitors provided measurements of the irregularities, as well as their size and speed. The dual-frequency GPS receiver

measured the total electron content of the ionosphere. If available, an all-sky imaging system measured two different thermosphere/ionosphere emissions from which the two-dimensional structure/motion of irregularities was observed. Those observations were used to calculate the density and height of the ionosphere. Two miniaturized Fabry-Perot interferometers measured the thermospheric neutral winds and temperatures. The two interferometers were separated by 300 km, allowing bistatic, common-volume measurements. Those measurements were useful for studying the response of the thermosphere to storms as well as for looking for the possible connection of gravity waves to the seeding of equatorial instabilities.

12. **South Atlantic Very Low Frequency Network**

    It was observed that the South Atlantic Very Low frequency Network (SAVNET) used the properties of very low frequency wave propagation on long distances between a transmitter and a receiver in the Earth-ionosphere waveguide. The waveguide was formed by the Earth's surface, which was an electrical conductor, and by the low ionosphere D-region at an altitude of approximately 70 km of altitude during diurnal conditions and the E-region at an altitude of approximately 90 km at night without the presence of solar radiation. The characteristics of very low frequency propagating waves (amplitude and phase velocity) in the waveguide critically depended on the geometry of the waveguide, the electrical conductivity of its borders and the geomagnetic field. All phenomena capable of changing those waveguide properties affected the characteristics of very low frequency propagation.

    SAVNET had two main objectives: (a) the indirect long-term monitoring of solar radiation; and (b) providing a diagnostic tool to study the ionosphere above the South Atlantic Magnetic Anomaly (SAMA) region during quiescent and geomagnetically disturbed periods. Further objectives of SAVNET were: (c) the study of ionospheric D-region properties during transient perturbations such as solar flares; (d) the diagnosis of extrasolar sources of ionospheric perturbations; (e) the observation of atmospheric phenomena producing ionospheric perturbations, like sprites, terrestrial gamma-ray flashes and seismo-electromagnetic processes; (f) the provision of experimental data sets to feed computational propagation codes in order to obtain daily templates of very-low-frequency wave properties on a given transmitter-receiver path; and (g) the study of peculiar properties of the ionosphere at high (southern) latitudes.

    The SAVNET base receiver was composed of two directional squared (3m x 3m) loop antennas and an isotopic vertical (6 m) antenna. The sensor signals were amplified and transported to an A/D audio card. The wave characteristics were provided by a Software Phase and Amplitude Logger computer code.

13. **Scintillation Network Decision Aid**

    It was recalled that SCINDA was a real-time, data-driven communication outage forecast and alert system. It aided in the specification and prediction of satellite communication degradation resulting from ionospheric scintillation in the equatorial region. Ionospheric disturbances caused rapid phase and amplitude fluctuations of satellite signals observed at or near the Earth's surface; those fluctuations were known as scintillation. The most intense natural scintillation events occurred during night-time hours within 20 degrees of the Earth's magnetic equator, a region encompassing more than one third of the Earth's surface. Scintillation affected radio signals up to a few GHz

frequencies and seriously degraded and disrupted satellite-based navigation and communication systems. SCINDA was designed to provide regional specification and short-term forecasts of scintillation activity to operational users in real time.

14. **Space Environmental Viewing and Analysis Network**

It was noted that the Space Environmental Viewing and Analysis Network (SEVAN) was an array of particle detectors located at middle and low latitudes which aimed to improve fundamental research of space weather conditions and to provide short- and long-term forecasts of the dangerous consequences of space storms. SEVAN detected changing fluxes of different species of secondary cosmic rays at different altitudes and latitudes, thus turning SEVAN into a powerful integrated device used to explore solar modulation effects.

# Annex I

## National and area coordinators of the International Space Weather Initiative

| Country or area | Coordinator | Affiliation |
|---|---|---|
| Algeria | N. Zaourar | Geophysical Laboratory, University of Sciences and Technology, Algiers |
| Argentina | C. Mandrini | Instituto de Astronomía y Física del Espacio, Buenos Aires |
| Armenia | A. Chilingarian | Cosmic Ray Division, Alikhanyan Physics Institute, Yerevan |
| Australia | B. Fraser | Centre for Space Physics, University of Newcastle |
| Austria | R. Nakamura | Institut für Weltraumforschung, Graz |
| Azerbaijan | E.S. Babayev | Shamakhy Astrophysical Observatory, Baku |
| Bahrain | M. Al Othman | Physics Department, Bahrain University |
| Belgium | G. Lapenta | Afdeling Plasma-astrofysica, Katholieke Universiteit Leuven |
| Benin | E. Houngninou | University of Abomey Calavi, Cotonou |
| Brazil | A. Dal Lago[a] | Instituto Nacional de Pesquisas Espaciais, Sao Paulo[a] |
| | J.P. Raulin[b] | Presbyterian Mackenzie University, Sao Paulo[b] |
| Bulgaria | K. Georgieva | Solar-Terrestrial Influences Laboratory, Sofia |
| Burkina Faso | F. Ouattara | University of Koudougou, Koudougou |
| Cameroon | E. Guemene Dountio | Ministry of Scientific Research and Innovation, Energy Research Laboratory |
| Canada | I. Mann | Department of Physics, University of Canada, Alberta |
| Cape Verde | J. Pimenta Lima | Instituto Nacional de Metereologica e Geofisica |
| China | W. Jing-Song | National Center for Space Weather, China Meteorological Administration |
| Congo | B. Dinga | Ministère de la recherche, Groupe de recherches en sciences exactes et naturelles, Brazzaville |
| Côte d'Ivoire | V. Doumbia | Laboratoire de physique de l'atmosphère, Université de Cocody, Abidjan |
| Czech Republic | F. Farnik[a] | Astronomical Institute, Ondřejov[a] |
| | L. Prech[b] | Department of Surface and Plasma Science, Faculty of Mathematics and Physics, Charles University, Prague[b] |
| Croatia | D. Roša | Zagreb Observatory |
| Democratic Republic of the Congo | B. Kahindo | Université de Kinshasa, Faculté Polytechnique, Kinshasa |
| Denmark | K. Galsgaard | The Niels Bohr Institute, Computational Astrophysics, Copenhagen |
| Ecuador | E. Lopez | Observatorio Astronómico de Quito, Interior del Parque La Alameda, Quito |
| Egypt | A. Mahrous | Space Weather Monitoring Center, Helwan |
| Ethiopia | B. Damtie | Department of Physics, Bahir Dar University |
| Finland | R. Vainio | Department of Physical Sciences, University of Helsinki |

| Country or area | Coordinator | Affiliation |
|---|---|---|
| France | N. Vilmer | Laboratoire d'études spatiales et d'instrumentation en astrophysique, Observatoire de Paris |
| Georgia | M.S. Gigolashvili | Abastumani Observatory |
| Germany | M. Danielides | Deutsches Zentrum für Luft- und Raumfahrt in der Helmholtz-Gemeinschaft |
| Greece | O. Malandraki | Institute for Astronomy and Astrophysics, Athens |
| Hungary | K. Kecskemety | Research Institute for Particle and Nuclear Physics, Budapest |
| India | P.K. Manoharan | Tata Institute of Fundamental Research, Radio Astronomy Centre |
| Indonesia | T. Djamaluddin[a] | National Institute of Aeronautics and Space, Bandung[a] |
| | D. Herdiwijaya[b] | Department of Astronomy, Institut Teknologi Bandung, Bandung[b] |
| Iraq | R. Al-Naimi | Department of Atmospheric Sciences, University of Baghdad |
| Ireland | P. Gallagher | School of Physics, Trinity College, Dublin |
| Israel | M. Gedalin | Department of Physics, Ben-Gurion University |
| Italy | M. Messerotti | Department of Physics, University of Trieste |
| Japan | T. Obara | Japan Aerospace Exploration Agency |
| Jordan | H. Sabat | Institute of Astronomy and Space Science, Al al-Bayt University, Mafraq |
| Kazakhstan | N. Makarenko | Institute of Mathematics, Almaty |
| Kenya | P. Baki | Department of Physics, University of Nairobi, Nairobi |
| Kuwait | I. Sabbah | Department of Physics, Faculty of Science, Kuwait University |
| Lebanon | R. Haijar | Department of Physics and Astronomy, Notre Dame University, Louaize |
| Libyan Arab Jamahiriya | A. Qader Abseim | Libyan Remote Sensing and Space Center |
| Malaysia | F. Bin Asillam | National Space Agency of Malaysia, Putrajaya |
| Mongolia | D. Batmunkh | Solar Physics Research Group, Mongolian Academy of Sciences |
| Morocco | N.-E. Najid | Université Hassan II Ain Chock, Faculté des Sciences Ain Chock, Casablanca |
| Nepal | J. Acharya | Mahendra Sanskrit University, Bakeemi Campus, Kathmandu |
| Niger | S. Madougou | Department of Physique, Ens University Abou Moumouni of Niamey |
| Nigeria | A.B. Rabiu | Department of Physics, Federal University of Technology, Akure, Ondo State |
| Norway | N. Ostgraard | Department of Physics and Technology, University of Bergen |
| Oman | S. Al-Shedhani | Physics Department, College of Science, Sultan Qaboos University, Al-Khoud |
| Peru | W. Guevara Day | University of Peru |
| Philippines | R. E.S. Otadoy | Department of Physics, University of San Carlos-Talamban Campus, Nasipit, Talamban, Cebu City |

| Country or area | Coordinator | Affiliation |
| --- | --- | --- |
| Poland | M. Tomczak | Astronomical Institute, University of Wroclaw, Wroclaw |
| Portugal | D. Maia | University of Lisbon |
| Puerto Rico | S. Gonzalez | Arecibo University, Arecibo |
| Qatar | S.S. Bin Jabor Althani | Astronomy Department, Qatar Science Club |
| Republic of Korea | Y.D. Park | Korea Astronomy and Space Science Institute, Daejeon |
| Romania | G. Maris | Institute of Geodynamics, Bucharest |
| Russian Federation | A. Stepanov[a] | Central Astronomical Observatory at Pulkovo, St. Petersburg[a] |
| | G.A Zherebtsov[b] | Institute of Solar-Terrestrial Physics, Russian Academy of Sciences, Siberian Branch, Irkutsk[b] |
| Rwanda | J. de Dieu Baziruwiha | Institut supérieur pedagogique, Kigali |
| Saudi Arabia | H. Basurah | Department of Astronomy, King Abdul Aziz University, Jeddah |
| Senegal | G. Sissoko | Groupe modelisation et simulation en energie solaire, Departement de physique, Universite Cheikh Anta Diop, Dakar |
| Serbia | I. Vince | Astronomical Observatory, Belgrade |
| Slovakia | I. Dorotovic | Slovak Central Observatory, Hurbanovo |
| South Africa | L.A. MacKinnel | Rhodes University, Grahamstown |
| Spain | J.R. Pacheco | Universidad de Alcalá |
| Sweden | H. Lundstedt | Swedish Institute of Space Physics, Lund |
| Switzerland | A. Csillaghy | University of Applied Sciences, Campus Brugg-Windisch |
| Thailand | B. Soonthornthum[a] | National Institute of Aeronautics and Space[a] |
| | D. Ruffolo[b] | Bandung Institute of Technology[b] |
| Tunisia | H. Ghalila | Laboratoire LSAMA, Départment de physique, Faculté des sciences de Tunis, Université de Tunis El Manar I |
| Turkey | A. Ozguc | Kandilli Observatory and E.R.I, Bogazici University, Istanbul |
| Ukraine | O. Litvinenko | Institute of Radio Astronomy NASU |
| United Arab Emirates | H.M.k. Al-Naimiy | United Arab Emirates University, Sharjah |
| United States | R. Smith | Geophysical Institute, University of Alaska |
| Uruguay | G. Tancredi | Observatorio Astronómico Los Molinos |
| Uzbekistan | S. Egamberdiev | Ulugbek Astronomical Institute |
| Viet Nam | H.T. Lan | Department of Atmosphere and Space Physics, Institute of Physics, Ho Chi Minh City |
| Yemen | A. Haq Sultan | Physics Department, Faculty of Science, Sanaa University |
| Zambia | N. Mwiinga | Department of Physics, School of Natural Sciences, University of Zambia, Lusaka |
| Palestine | I. Barghouthi | Department of Physics, Faculty of Science, Al-Quds University, Jerusalem |
| Taiwan Province of China | C.Z.F. Cheng | Plasma and Space Science Center, Tainan |

[a] Primary contact.
[b] Secondary contact.

# Annex II

## International Space Weather Initiative instrument distribution by country or area

| Country or area | Number of instruments | Type of instrument(s) |
|---|---|---|
| Algeria | 7 | AMBER (1), AWESOME (1), CHAIN (1), GPS-Africa (1), MAG-Africa (1), SID (2) |
| Antarctica | 2 | AWESOME (1), SID (1) |
| Argentina | 1 | SAVNET (1) |
| Armenia | 1 | SEVAN (1) |
| Australia | 14 | CALLISTO (2), GMDN (1), MAGDAS (10), OMTI (1) |
| Austria | 2 | AWESOME (1), SID (1) |
| Azerbaijan | 3 | AWESOME (1), SID (2) |
| Belgium | 1 | CALLISTO (1) |
| Benin | 1 | GPS-Africa (1) |
| Bosnia and Herzegovina | 1 | SID (1) |
| Botswana | 1 | GPS-Africa (1) |
| Brazil | 16 | CALLISTO (1), GMDN (1), MAGDAS (2), RENOIR (2), SAVNET (4), SCINDA (3), SID (3) |
| Bulgaria | 3 | SEVAN (1), SID (2) |
| Burkina Faso | 3 | GPS-Africa (2), SID (1) |
| Cameroon | 2 | AMBER (1), SCINDA (1) |
| Canada | 10 | MAGDAS (1), OMTI (2), SID (7) |
| Cape Verde | 1 | GPS-Africa (1) |
| Central African Republic | 1 | MAG-Africa (1) |
| Chile | 2 | SCINDA (1), SID (1) |
| China | 10 | SID (9), SEVAN (1) |
| Colombia | 3 | SCINDA (1), SID (2) |
| Cong | 4 | SCINDA (1), SID (3) |
| Costa Rica | 2 | CALLISTO (1), SEVAN (1) |
| Côte d'Ivoire | 4 | MAGDAS (1), MAG-Africa (2), SCINDA (1) |
| Croatia | 2 | SEVAN (1), SID (1) |
| Cyprus | 1 | AWESOME (1) |
| Czech Republic | 2 | CALLISTO (1), SID (1) |
| Democratic Republic of the Congo | 2 | SID (2) |
| Ecuador | 1 | AWESOME (1) |
| Egypt | 7 | AWESOME (1), CALLISTO (1), CIDR (1), MAGDAS (2), SID (2) |
| Ethiopia | 11 | AMBER (1), AWESOME (1), MAGDAS (1), MAG-Africa (1), SCINDA (2), SID (5) |
| Fiji | 1 | AWESOME (1) |

| Country or area | Number of instruments | Type of instrument(s) |
|---|---|---|
| Finland | 1 | CALLISTO (1) |
| France | 4 | SID (4) |
| Gabon | 2 | GPS-Africa (2) |
| Germany | 21 | CALLISTO (1), SEVAN (1), SID (19) |
| Ghana | 1 | GPS-Africa (1) |
| Greece | 2 | AWESOME (1), SID (1) |
| Guyana | 1 | SID (1) |
| India | 19 | AWESOME (4), CALLISTO (2), MAGDAS (1), SEVAN (1), SID (11) |
| Indonesia | 5 | MAGDAS (3), SEVAN (1), SID (1) |
| Ireland | 8 | AWESOME (1), CALLISTO (1), SID (6) |
| Israel | 2 | AWESOME (1), SEVAN (1) |
| Italy | 32 | MAGDAS (1), SID (31) |
| Japan | 12 | CHAIN (1), GMDN (1), MAGDAS (6), OMTI (4) |
| Jordan | 1 | AWESOME (1) |
| Kenya | 6 | GPS-Africa (1), MAGDAS (1), SCINDA (1), SID (3) |
| Kuwait | 1 | GMDN (1) |
| Lebanon | 6 | SID (6) |
| Libyan Arab Jamahiriya | 2 | AWESOME (1), SID (1) |
| Madagascar | 1 | MAG-Africa (1) |
| Malaysia | 3 | AWESOME (1), MAGDAS (1), OMTI (1) |
| Mali | 4 | GPS-Africa (2), MAG-Africa (2) |
| Mauritius | 1 | CALLISTO (1) |
| Mexico | 5 | CALLISTO (1), SID (4) |
| Micronesia, Federated States of | 1 | MAGDAS (1) |
| Mongolia | 12 | AWESOME (1), CALLISTO (1), SID (10) |
| Morocco | 2 | AWESOME (1), GPS-Africa (1) |
| Mozambique | 3 | GPS-Africa (1), MAGDAS (1), SID (1) |
| Namibia | 4 | AMBER (1), GPS-Africa (1), MAG-Africa (1), SID (1) |
| Netherlands | 1 | SID (1) |
| New Zealand | 3 | SID (3) |
| Niger | 1 | GPS-Africa (1) |
| Nigeria | 32 | AMBER (1), MAGDAS (3), SCINDA (2), SID (26) |
| Norway | 1 | OMTI (1) |
| Peru | 8 | AWESOME (1), CHAIN (1), CIDR (1), MAGDAS (1), SAVNET (3), SCINDA (1) |
| Philippines | 7 | MAGDAS (6), SCINDA (1) |
| Poland | 1 | AWESOME (1) |
| Portugal | 1 | SID (1) |
| Republic of Korea | 2 | SID (1), CALLISTO (1) |

| Country or area | Number of instruments | Type of instrument(s) |
| --- | --- | --- |
| Romania | 2 | SID (2) |
| Russian Federation | 6 | CALLISTO (1), MAGDAS (3), OMTI (2) |
| Sao Tome and Principe | 1 | GPS-Africa (1) |
| Saudi Arabia | 2 | AWESOME (1), SCINDA (1) |
| Senegal | 3 | GPS-Africa (1), MAG-Africa (1), SID (1) |
| Serbia | 2 | AWESOME (1), SID (1) |
| Slovakia | 2 | SEVAN (1), SID (1) |
| South Africa | 20 | GPS-Africa (7), MAGDAS (2), MAG-Africa (2), SID (9) |
| Spain | 1 | MAG-Africa (1) |
| Sri Lanka | 1 | SID (1) |
| Sudan | 1 | MAGDAS (1) |
| Switzerland | 4 | CALLISTO (3), SID (1) |
| Thailand | 4 | OMTI (1), SID (3) |
| Tunisia | 4 | AWESOME (1), SID (3) |
| Turkey | 3 | AWESOME (1), SID (2) |
| United Arab Emirates | 1 | AWESOME (1) |
| United Kingdom of Great Britain and Northern Ireland | 8 | MAG-Africa (1), SID (7) |
| United Republic of Tanzania | 2 | GPS-Africa (1), MAGDAS (1) |
| United States of America | 172 | AWESOME (2), CALLISTO (1), CIDR (6), MAGDAS (2), OMTI (1), SID (160) |
| Uganda | 3 | GPS-Africa (1), SID (2) |
| Uruguay | 3 | SID (3) |
| Uzbekistan | 2 | AWESOME (1), SID (1) |
| Viet Nam | 2 | AWESOME (1), MAGDAS (1) |
| Zambia | 4 | GPS-Africa (1), MAGDAS (1), SID (2) |
| Taiwan Province of China | 1 | MAGDAS (1) |